\documentclass[aps,prb,twocolumn,floatfix,showpacs,showkeys,amssymb,amsmath]{revtex4}
\usepackage{graphicx}
\usepackage{amsmath}
\usepackage{amsfonts}
\usepackage{amssymb}

\begin{document}
\title{Coherent State Monitoring in Quantum Dots}
\author{Ameenah N. Al-Ahmadi}

\author{Sergio E. Ulloa }
\affiliation{Department of Physics and Astronomy, and Nanoscale
and Quantum Phenomena Institute, Ohio University, Athens, OH
45701-2979 }

\date{\today } 

\begin{abstract}
We study the dynamics of excitonic states in dimer and trimer
arrangements of colloidal quantum dots using a density matrix
approach. The dots are coupled via a dipole-dipole interaction
akin to the F\"orster mechanism. Coherent oscillations of tuned
donor dots are shown to appear as plateaus in the acceptor dot,
and therefore in its optical response. This behavior provides one
with an interesting and unique handle to monitor the quantum state
of the dimer, an ``eavesdroping arrangement."  A trimer cluster in
a symmetrical loop shows steady states in a shorter characteristic
time than the typical radiative lifetime of the dots. Breaking the
symmetry of the loop results again in damping oscillatory states
in the donor dots and plateaus in the eavesdropping/acceptor dot.
The use of realistic parameters allows direct comparison with
recent experiments and indicates that coherent state monitoring is
possible in real experiments.
\end{abstract}
\maketitle

The individual optical and electronic properties of quantum dots
(QDs) can be controlled by their size, shape, and
composition.\cite{Efros,BigRev} Optical studies on different QD
systems, such as CdSe, \cite{Bawendi1,Klimov1} CdS, \cite{Weller}
and InP, \cite{Nozik} have revealed information about the coupling
between collections of dots.  This coupling may take place via
direct charge transfer (tunneling), and/or via long-range energy
transfer or F\"orster interaction, in which excited donor ($D$)
dots transfer their energy to unexcited acceptor ($A$) dots.
F\"orster developed the theoretical treatment of energy transfer
in organic molecules,\cite{Forster} and is now routinely observed
even in molecular self-assembled layers.\cite{Ana} This theory
assumes that the energy transfer arises from dipole-dipole
interaction. Although higher multipolar interactions are possible
in principle, they are negligible in typical systems,\cite{jpns}
but important in others, such as closely packed metal
nanoparticles. \cite{Claro} F\"orster derived an expression for
the resonant energy transfer rate between donors and acceptors, $
K_{\textit{DA}}  = ({2\pi }/ \hbar )V_F^2 \, \Theta \,$,
where
\begin{equation}
V_F  = {{\mu _D \mu _A } \over {\varepsilon  R^3 }}\kappa {\rm }
\, . \label{t2}
\end{equation}
Here, $\mu _D$  ($\mu _A)$ is the dipole moment of the donor
(acceptor), $R$ is the separation between $D$ and $A$ centers,
$\varepsilon$ is the dielectric constant of the medium.  $\kappa$
is an angular dipole orientation factor; assuming the dipole
orientations for $D$ and $A$ to be (anti- or) parallel, we set
$\kappa \simeq 1$.  $\Theta$ is the spectral overlap integral
between normalized donor emission and acceptor absorption
lineshapes. Energy transfer between QDs has been verified in
beautiful experiments by different groups,
\cite{Bawendi1,Klimov1,Klimov2,Awschalom} opening a number of
interesting possibilities.  For example, one can consider the
low-temperature non-dissipative coupling between dots and the
corresponding coherent oscillations describing excitation
transfer. This provides one with a new tool to control the
coherent state of an optical excitation in the QDs, controllably
monitor coherent spin transfer between dots,\cite{ Awschalom,
Govorov} and possibly new physical implementations of quantum
computation concepts.

In this paper we use a density matrix approach to study the
dynamics of exciton states in dimer and trimer arrangements of
colloidal quantum dots. We consider that each quantum dot has two
main exciton states, one optically passive (dark) and another
active (bright), to account for the well-known symmetries in II-VI
nanocrystals. \cite{Efros}  The dots are assumed to be in close
proximity, thanks to molecular linkers or spacers that allow
dipolar coupling but yet prevent direct carrier hopping. We
analyze the time evolution of each exciton state after different
pumping pulses, and for different structural parameters.  We find,
for example, that at low temperatures and for realistically
attainable systems, one could monitor the coherent oscillations
between neighboring (and nearly identical) dots using an
``eavesdropping" acceptor dot nearby. The monitoring acceptor dot
is shown to exhibit periodic photoluminescence (or absorption)
plateaus in the sub-nanosecond regime, and only weakly (although
irremediably) affecting the coherent oscillations in the
sympathetic dimer. We also explore other geometries and regimes
and show how one can exploit the flexibility in dot cluster
features to probe the quantum mechanical states of these systems.

{\em Theory and Model}. We use a Markovian equation to describe
the dynamics of the exciton states in the QD system
\begin{equation}
\frac{\partial\hat{\rho}_{ij}}{\partial t}=
-\frac{i}{\hbar}\langle
i|[\hat{H},\hat{\rho}|j\rangle-\sum_{lk}\Gamma_{lk,ij}\hat{\rho}_{lk}
,\label{t3}
\end{equation}
where $\hat{H}$ and $\hat{\rho}$ are  the Hamiltonian and density
operators, respectively, and $\Gamma_{lk,ij}$ is the relaxation
matrix. \cite{orange book} The Hamiltonian that describes the
system is the excitonic Hubbard model given by\cite{Bryant}
\begin{equation}
H = \sum\limits_i^N {U_i } c_i^{{\rm \dag } } c_i d_i^{{\rm \dag }
} d_i  + \sum\limits_{i \ne j}^N {V_{Fi_n j_m } c_{in}^{\rm \dag }
d_{in}^{\rm \dag } d_{jm} c_{jm} } ,\label{t4}
\end{equation}
where $U_i$ is the binding energy of the exciton in the dot,
$c^{{\rm \dag }}$ and $d^{{\rm \dag }}$ are electron and hole
operators, and $V_{Fi_n j_m }$ is the coupling constant that
governs the exciton energy transfer between level $n$ of dot $i$
and level $m$ of dot $j$. The coupling constant in principle
includes all possible mechanisms that allow the energy transfer to
take place. This Hamiltonian can describe, for instance (with the
possible addition of phonon degrees of freedom), how the energy
moves among donors before emission occurs (exciton migration),
\cite{Bartolo} or how the excitation transfers irreversibly from
donor to acceptor and is then emitted out of the system as a real
photon (F\"orster transfer). \cite{Forster}  We will focus here on
the coherent coupling between dots (a low temperature regime). The
dipole moments entering $V_{Fi_n j_m} $ are modeled by a hard wall
confinement potential, proven successful in the description of
colloidal quantum dots. \cite{Efros}  We correct the
overestimation of the exciton energy ground state for small sizes
by rescaling the gap to match the experimental
results.\cite{Bawendi2} For simplicity, the dot is assumed to have
two exciton levels.  The high-energy bright level (absorbing line)
has a rapid relaxation rate $(\lesssim 1$ ps at $\sim 10$ K) to
the dark (low-energy) level (emitting line) as known from
experiments [see inset (b) of Fig.\ 1]. The lifetime of the dark
level is much longer ($\sim 20$ ns) than the time scale of our
calculations ($\sim 1$ ns). The linewidth of the dark states
becomes then negligible in the time window of interest. We should
mention that inclusion of a more detailed level description of the
dots is straightforward, although not essential for our
conclusions, as we will see.  The experimental values obtained
from the luminescence data of the exciton in CdSe nanocrystals are
used to estimate the coupling constants. \cite{Bawendi3}

{\em Dimer}. Consider first the dynamics of a \textit{dimer}
consisting of a donor and acceptor dot pair (such as the $D_2$-$A$
pair in the insets in Fig.\ 1).  The system may be reduced to
three main levels, neglecting the bright exciton level of the
donor dot because of its characteristic fast relaxation time. The
exact analytical solution of such a three-level system shows that
there is an effective decay of the donor lower level which depends
on the coupling $V_{F}$ (itself a function of dot separation and
sizes), the width of the bright exciton state in the acceptor dot
$\Gamma_{A}$, and on the detuning $\omega$ between the donor dark
state and the acceptor bright state. The effective relaxation for
weak coupling $V_F \ll \Gamma_{A}$, is given by the rate
\begin{equation}
\Gamma _{\textit{eff}}  = {{V_F^2 \, \Gamma_{A} } \over {\left(
{{\Gamma_{A} / 2}} \right)^2  + \omega ^2 }} \, , \label{t5}
\end{equation}
while for $V_F \gg \Gamma_A$, $\Gamma_{\textit{eff}} \simeq
\Gamma_A /2$. The linewidth $\Gamma_A$ of the exciton levels in a
single QD is of course a function of temperature,\cite{Klimov3}
and QD radius.\cite{Bawendi3}  These equations reveal that a dimer
can be thought of and used as a {\em tunable linewidth} ``level"
by controlling the coupling $V_F$, either by changing the distance
between donor and acceptor dots or by changing their sizes (which
changes $\omega$ and weakly $\Gamma_A$).  Notice that the dark
level in the donor dot transfers energy to the acceptor bright
state as a {\em virtual} (non-radiative) process. \cite{footnote}

\begin{figure}[tbp]
\includegraphics*[width=1.0\linewidth]{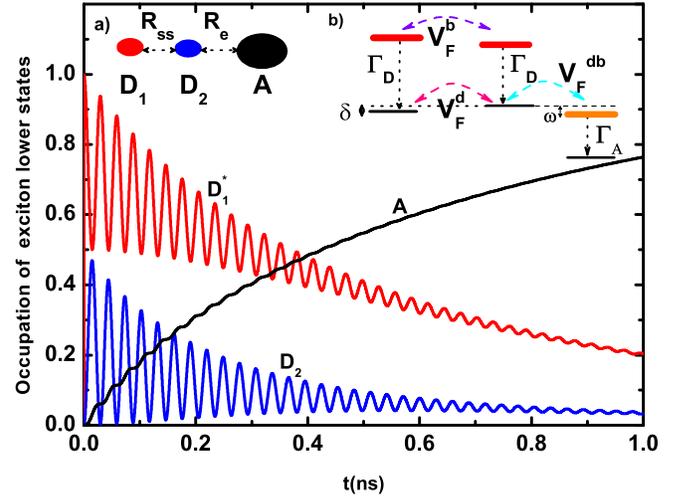}
\caption{Time evolution of occupation probability of lower exciton
state in each dot. Parameters used are $V_F^d = - 0.05$ meV, $
V_F^b  =  - 1$  meV, $ V_F^{db}  =  - 0.21$ meV, $ \Gamma _D  =
32$ meV, $ \Gamma _A  = 30.5$  meV, $\delta = \omega = 0.1$ meV,
and $R_{\textit{ss}}=R_e= 11$\AA.  Inset (a) shows diagram of
trimer chain and (b) corresponding energy diagram. Donor dot $D_1$
has been excited at $t=0$.  Notice plateaus in occupation of dot
A.} \label{fig1}
\end{figure}

\begin{figure}[tbp]
\includegraphics*[width=1.0\linewidth]{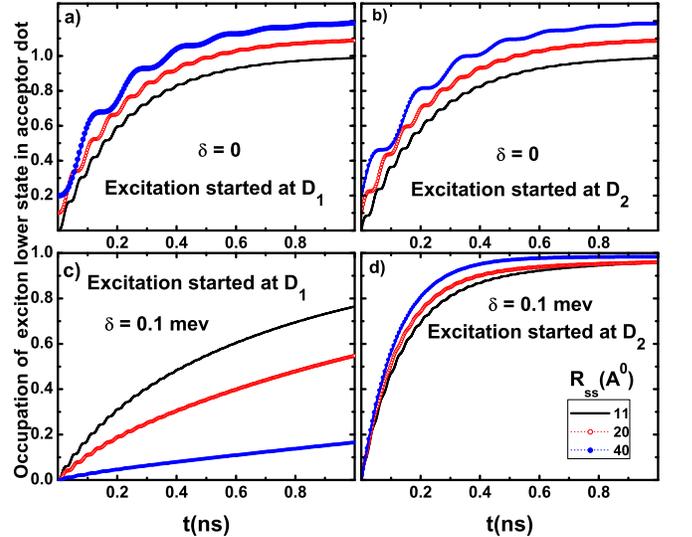}
\caption{Dynamics of exciton lower state occupation of acceptor
dot for different $R_{ss}$ values ($R_e = 11$\AA \ fixed as in
Fig.\ 1). (a) Initial excitation (pumping) of $D_1$ for $\delta
=\omega=0$. (b) Pumping $D_2$, $\delta =0$. (c) Pumping $D_1$,
$\delta = 0.1$ meV.\@ (d) Pumping $D_2$, $\delta = 0.1$ meV.\@
Curves in (a) and (b) are offset vertically for clarity.  Notice
large variation in characteristic growth times, as well as
plateaus for nearly all conditions.} \label{fig2}
\end{figure}

{\em Linear Trimer}. Let us consider two donor dots coupled to a
third acceptor dot, forming a trimer.  Figure 1 shows the
oscillations in occupation of the different exciton \textit{lower}
states in each dot after the first donor dot ($D_1$) is resonantly
excited. All six levels in the system have been used in the
numerical solution of the density matrix evolution. The red and
blue curves represent the low energy exciton level for the first
and second donor dot, respectively.  The black curve is for the
acceptor lower state. It is clear that the coupling between the
donor lower states $V_{F}^{d}$, induces coherent oscillations of
the exciton population in the donor pair, despite the small
detuning $\delta$ between the low energy donor states. These
oscillations show a period $\tau_{F}^{d}\backsimeq h
/\sqrt{\delta^2 + 4{V_F^d}^2} = 29$ ps, with an overall slow
decay, also seen as growth in the acceptor dark state, with a rise
time $\backsimeq 0.7$ ns. Most importantly, the acceptor is shown
to act as a nearly ideal ``eavesdropping" point that can
effectively monitor coherent oscillations between the two donor
dots without totally collapsing them: The coherent oscillations
between the nearly tuned donors appear as plateaus in the acceptor
dark level population with the same time period of 29 ps.  These
plateaus persist over the first 0.3 ns, and could possibly be
monitored by time-resolved differential absorption measurements of
the acceptor state.

\begin{figure}[tbp]
\includegraphics*[width=1.0\linewidth]{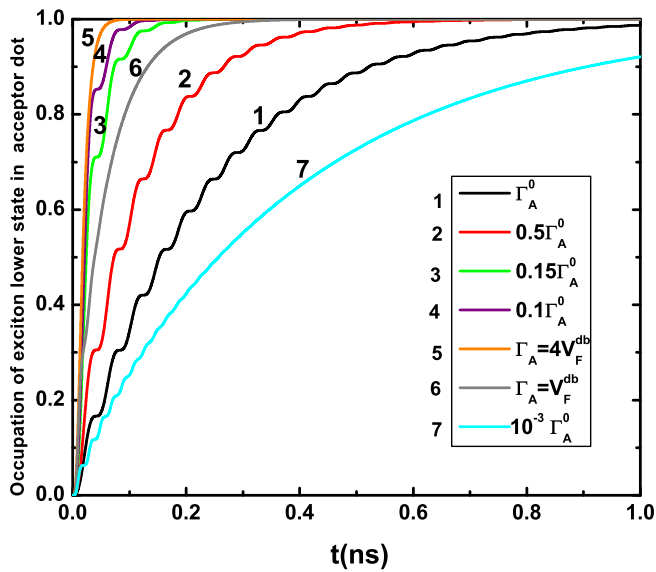}
\caption{Dynamics of exciton lower state population at the
acceptor dot with varying $\Gamma_A$. $\Gamma_A^0 = 30.5$ meV as
in Figs.\ 1 and 2. Notice absence of oscillation plateaus and
fastest rise time for $\Gamma_A = 4V_F^{db}$.} \label{fig3}
\end{figure}

The effects of interdot separation on the dynamics of the lower
acceptor dot state are shown in Fig.\ 2 for different pumping
conditions. Increasing the separation $R_{\textit{ss}}$ between
the surfaces of the two donors, $D_1$ and $D_2$ in Fig.\ 1,
reduces the coupling constant as $(a_1+a_2+R_{ss})^{-3} $, where
$a_1$ ($a_2$) is the radius of $D_1$ ($D_2$).  The oscillation
period is longer for larger $R_{\textit{ss}}$ since $V_F^d$
becomes smaller, and fewer plateaus appear in the low energy state
of $A$ over the same time interval. Notice that the $D_2$-$A$
distance in the trimer remains constant throughout Fig.\ 2; larger
$R_e$ separations would result in less defined plateaus in the
growth curves in Fig.\ 2. The behavior is slightly different when
we start the excitation in either the first or second donor, $D_1$
or $D_2$, as shown in Fig.\ 2a and 2b. In these cases the growth
rate remains nearly constant ($\simeq \Gamma /2$ or 0.24 ns, with
$\Gamma$ as in Eq.\ \ref{t5}), as it is related to the coupling
between $D_2$ and $A$ kept fixed, while the oscillation period
changes with $R_{ss}$. Notice also that there is a delay time for
the appearance of the first plateau in 2a which is different in
each case. This delay time, $\sim 13$ ps, corresponds to the
accumulation time of the amplitude occupying the lower exciton
level in $D_1$. Panels (c) and (d) of the figure represent two
donors with a slightly different size and corresponding detuning
$\delta = \omega =0.1$ meV. Notice that a small detuning allows
one to spectrally select which of the two donor dots is actually
pumped in experiments, so that different initial conditions can be
tested.  The occupation of the lower exciton state in the acceptor
dot exhibits here {\em different} overall slopes that rapidly
increase with decreasing $R_{ss}$, as shown in Fig.\ 2c in the
case of $D_1$ pumping. It is clear that weaker coupling (larger
$R_{\textit{ss}}$) between $D_1$-$D_2$ results in slower building
up of the amplitude in $A$, as one would expect. Moreover, the
number and amplitude of the plateaus decrease for larger
$R_{\textit{ss}}$. In contrast, the dynamics in 2d, after exciting
$D_2$, shows minor changes in slope (given by Eq.\ \ref{t5},
$\simeq 0.12$ ns), and also clear disappearance of the plateaus as
the distance $R_{ss}$ increases. This behavior can be understood
by analogy to Rabi oscillations, where the Rabi period must be
shorter than the effective decay given by the homogeneous
linewidth $\Gamma_{\textit{eff}}$. The detuning increases the Rabi
frequency, but slows the growth rate of occupation of the acceptor
lower state. The strong $R_\textit{{ss}}$ dependence shown here
emphasizes the need for close and well-controlled proximity of the
$D_1$-$D_2$ separation. As experiments in this field show
outstanding control, one would expect that this requirement is
easily fulfilled.

\begin{figure}[tbp]
\includegraphics*[width=1.0\linewidth]{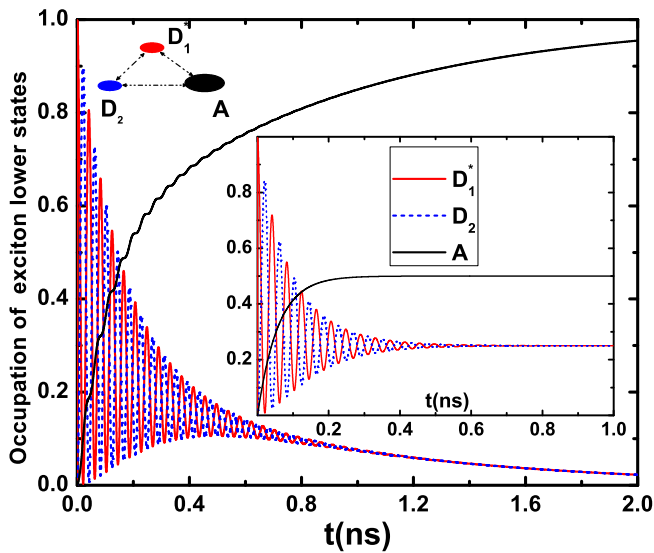}
\caption{Evolution of lower exciton state at each dot arranged in
an asymmetrical cluster loop as shown in top left inset.
Oscillation plateaus are visible in eavesdropping $A$ dot. Central
inset shows results for a {\em symmetrical} loop; notice absence
of plateaus in the $A$-response and saturation at
$\frac{1}{4}$-filling in this time window.} \label{fig4}
\end{figure}

In Fig.\ 3 we plot the probability of finding the exciton in the
acceptor dot in the trimer for different values of the linewidth
of the acceptor bright state. The occupation of the acceptor
low-energy state grows exponentially with time, and the effective
growth rate increases with decreasing linewidth $\Gamma_A$. It is
interesting that decreasing $\Gamma_A$ results in a {\em total
suppression} of the coherent plateaus at a special value $\Gamma
\simeq 4V_F^{db} \simeq 1$ meV (more on these and related values
below).  Further decreasing $\Gamma_A$ results in a smaller growth
constant and the plateaus becoming visible again. For a
description of this behavior we divide the relative values of
$\Gamma_A$ and $V_F^{db}$ into three regions: $\Gamma_A \gg
V_F^{db}$, $\Gamma_A \sim 4V_F^{db}$, and $\Gamma_A \ll V_F^{db}$.
In the first region, where $V_F^d ,V_F^{db} \ll \Gamma_A$, the
analytical solution of the dimer may be employed to reduce the
second donor and acceptor to a two-level system having an
effective damping given by Eq.\ (\ref{t5}). Therefore the
$D_1$-$D_2$ coherent oscillation frequency $V_F^d$ faces this
effective damping, as in the case depicted in Figs.\ 1 and 2. In
the second region, where $\Gamma_A \sim 4V_F^{db}$, the
oscillation and plateaus are suppressed, as the effective
oscillation frequency is vanishingly small, and only a smooth
damping persists. In the last region, where $\Gamma_A \ll
V_F^{db}$, the coherent oscillations reoccur and the oscillation
survives a damping rate equal to $\Gamma_A/2$. From this result
and those above, it is clear that changing the system parameters
changes the effective damping of the excitation in the donor.
Doing this, one can control the relative coupling between donors
and the effective damping of the nearby dimer. For example, the
coupling constant of the two lower states in the donor is
$V_F^{d}=-0.05$ meV and the effective damping for our dimer is 5.8
$\mu$eV, resulting from a F\"orster coupling of 0.21 meV for
$R_{e} = 11$ \AA  \ between donor and acceptor dots. Making the
$D_2$-$A$ dot surfaces touch we can increase the coupling to
$\sim$ 0.7 meV, which gives an effective damping of $\sim$ 0.06
meV and hence we can reduce the interaction $V_F^{d}$ to a value
less than the effective damping.  In this example, the interval
above and below the resonant condition $\Gamma_A \sim 4V_F^{db}$
can be covered in experiments.

{\em Trimer Cluster}. In Fig.\ 4 we explore the arrangement of a
trimer arranged as a cluster ``loop," with two identical donor
dots and an acceptor that has a bright exciton level in resonance
with the lower exciton level in the donors and is coupled to both
dots. The central inset shows results for the symmetric case in
which the distances between dots are identical. This symmetrical
loop shows essentially {\em steady} states of the donor and
acceptor dots in a characteristic time that is less than the
typical radiative time of the dots. This surprising lack of
oscillation plateaus in $A$, as well as the long-lasting plateau
at quarter-filling in $D_1$ and $D_2$ could be a way to test for
the symmetry of a cluster sample in experiments. The cancellation
of the oscillation plateaus is a natural result of the in-phase
(simultaneous) excitation of the $D$ dots and a direct proof of
coherent behavior.  On the other hand, breaking the symmetry of
the loop, as shown in the main curves in Fig.\ 4, restores damping
to the oscillatory behavior in the donor dots and plateaus to the
eavesdropping/acceptor dot. The characteristic oscillation period
and plateau rising times are similar to the linear arrangements.

{\em Conclusions}. We have used the density matrix approach to
investigate the dynamics of the exciton lower states in quantum
dot clusters arranged linearly and in triangular loops. The dots
are coupled via dipole-dipole excitonic interactions, and the
energy transfer processes can take place among donor dots, so that
the excitation energy may reach an acceptor dot after ``hopping"
(near) resonantly among donors. Coherent oscillations are induced
in donors and they appear as plateaus in the acceptor dot.  The
acceptor then behaves as a nearly ideal ``eavesdropping" observer
point that can effectively monitor the oscillations without
strongly affecting them. Although our results here are based on a
simplified model of the dot, it is clear that the complexity of
multilevel dynamics would contribute to increase $\Gamma_A$ rates
slightly but would not affect our main results. We believe this
phenomenon allows the possibility of monitoring states where
perhaps the spin of the carriers could be effectively initialized
via circularly polarized pumping.  A description of this and
related regimes will be presented elsewhere. 

This work was supported by the Indiana 21$^{st}$ Century Fund and
the Condensed Matter and Surface Sciences Program at OU. We thank
helpful discussions with G. Bryant, J.M. Villas-B\^{o}as, G.P. Van
Patten and the Ball State University QD team.


\begin{thebibliography}{99}
\bibitem {footnote} We should remark that it is well known that
the dark lower energy exciton state is not optically active due to
its angular momentum $F = \pm 2$. \cite{Efros} However, this state
does have non-zero dipole moment which can then produce a
F\"{o}rster coupling and its energy be (near) resonantly
transfered to neighboring dots via virtual non-radiative
intermediate states.
\end{thebibliography}
\end{document}